\documentclass[10pt, conference, compsocconf]{IEEEtran}

\usepackage{courier}
\usepackage{amsmath}
\usepackage{amssymb}
\usepackage{amsfonts}
\usepackage{xcolor}
\usepackage{multirow}
\usepackage{float}
\usepackage{graphicx}
\usepackage{graphics}
\usepackage{subfigure}
\usepackage{float}
\usepackage{bigstrut}
\usepackage{booktabs}
\usepackage{bibentry}
\usepackage{url}

\begin{document}

\title{Context-aware Sequential Recommendation}

\author{
\IEEEauthorblockN{Qiang Liu$^{1,2}$, Shu Wu$^1$, Diyi Wang$^3$, Zhaokang Li$^4$, Liang Wang$^{1,2}$}
\IEEEauthorblockA{$^1$Institute of Automation, Chinese Academy of Sciences\\
$^2$University of Chinese Academy of Sciences\\
$^3$Northeastern University $^4$Rice University\\
\{qiang.liu, shu.wu\}@nlpr.ia.ac.cn, wang.di@husky.neu.edu,\\
 zhaokang.li@rice.edu, wangliang@nlpr.ia.ac.cn}
}
\maketitle

\begin{abstract}
Since sequential information plays an important role in modeling user behaviors, various sequential recommendation methods have been proposed. Methods based on Markov assumption are widely-used, but independently combine several most recent components. Recently, Recurrent Neural Networks (RNN) based methods have been successfully applied in several sequential modeling tasks. However, for real-world applications, these methods have difficulty in modeling the contextual information, which has been proved to be very important for behavior modeling. In this paper, we propose a novel model, named Context-Aware Recurrent Neural Networks (CA-RNN). Instead of using the constant input matrix and transition matrix in conventional RNN models, CA-RNN employs adaptive context-specific input matrices and adaptive context-specific transition matrices. The adaptive context-specific input matrices capture external situations where user behaviors happen, such as time, location, weather and so on. And the adaptive context-specific transition matrices capture how lengths of time intervals between adjacent behaviors in historical sequences affect the transition of global sequential features. Experimental results show that the proposed CA-RNN model yields significant improvements over state-of-the-art sequential recommendation methods and context-aware recommendation methods on two public datasets, i.e., the Taobao dataset and the Movielens-1M dataset.
\end{abstract}

\IEEEpeerreviewmaketitle

\section{Introduction}
Nowadays, people are overwhelmed by a huge amount of information. Recommender systems have been an important tool for users to filter information and locate their preference. Since historical behaviors in different time periods have different effects on users' behaviors, the importance of sequential information in recommender systems has been gradually recognized by researchers. Methods based on Markov assumption, including Factorizing Personalized Markov Chain (FPMC) \cite{rendle2010factorizing} and Hierarchical Representation Model (HRM) \cite{wang2015learning}, have been widely used for sequential prediction. However, a major problem of these methods is that these methods independently combine several most recent components. To resolve this deficiency, Recurrent Neural Networks (RNN) have been employed to model global sequential dependency among all possible components. It achieves the state-of-the-art performance in different applications, e.g., sentence modeling \cite{mikolov2010recurrent}, click prediction \cite{zhang2014sequential}, location prediction \cite{liu2016strnn}, and next basket recommendation \cite{yu2016dream}.

\begin{figure*}[htb]
\centering
\setlength{\abovecaptionskip}{0pt}
\setlength{\belowcaptionskip}{-15pt}
\includegraphics[width=1\linewidth]{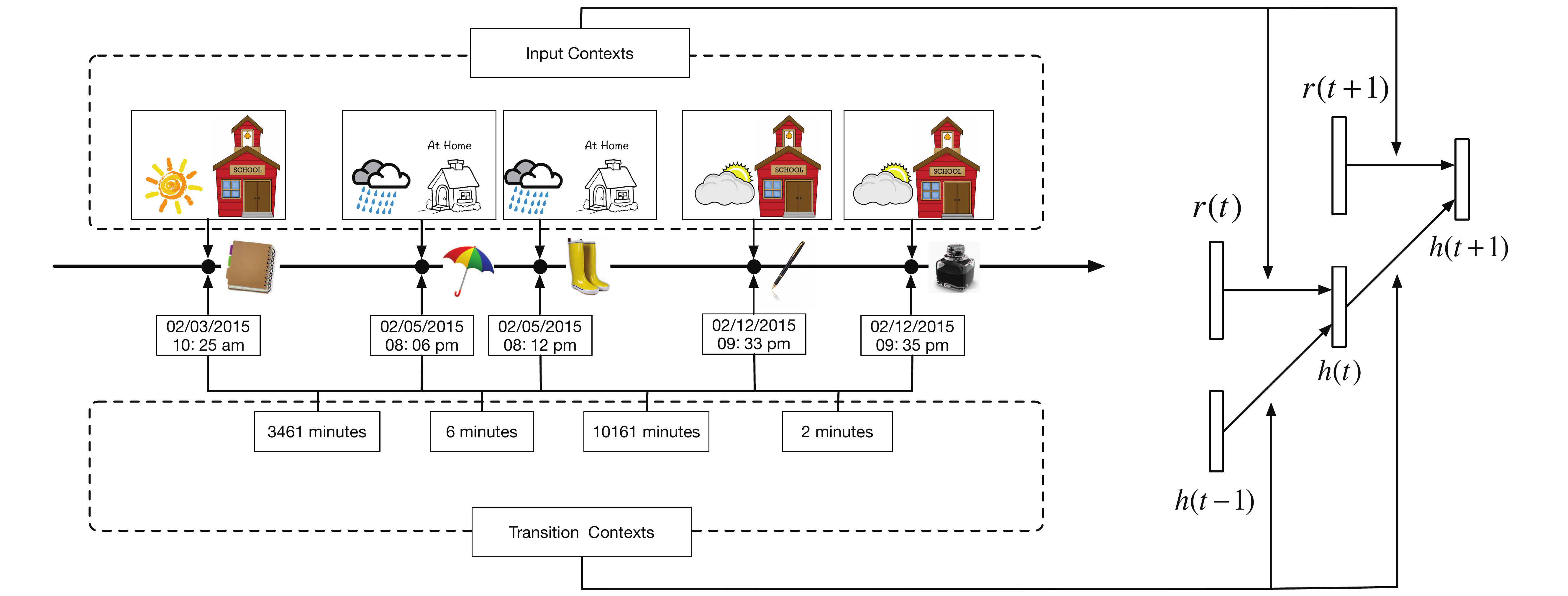}
\caption{The purchasing sequence of a user as an example of context-aware sequential recommendation. The left part shows input and transition contexts in a behavioral sequence. Input contexts mean external situations that users conduct behaviors, and transition contexts denote time intervals between adjacent behaviors. The right part illustrates how input and transition contexts contribute to predicting a user's next behavior in recurrent neural networks.}
\label{fig:Model-intro}
\end{figure*}

With enhanced ability in collecting information, a great amount of contextual information, such as location, time, weather and so on, can be collected by systems. Besides, the contextual information has been proved to be useful in determining users' preferences in recommender systems \cite{adomavicius2011context}. In real scenarios, as shown in Figure \ref{fig:Model-intro}, applications usually contain not only sequential information but also a large amount of contextual information. Though RNN and other sequential recommendation methods have achieved satisfactory performances in sequential prediction, they still have difficulty in modeling rich contextual information in real scenarios. On the other hand, context-aware recommendation has been extensively studied, and several methods have been proposed recently, such as Factorization Machine (FM) \cite{rendle2011fast}, Tensor Factorization for MAP maximization (TFMAP) model \cite{shi2012tfmap}, CARS2 \cite{shi2014cars} and Contextual Operating Tensor (COT) \cite{liu2015cot}\cite{wu2016contextual}. But these context-aware recommender methods can not take sequential information into consideration.

To construct a model to capture the sequential information and contextual information simultaneously, we first investigate the properties of sequential behavioral histories in real scenarios. Here, we conclude two types of contexts, i.e., \textbf{input contexts} and \textbf{transition contexts}. Input contexts denote external situations under which input elements occur in behavioral sequences, that is to say, input contexts are external contexts under which users conduct behaviors. Such contexts usually include location (home or working place), time (weekdays or weekends, morning or evening), weather (sunny or rainy), etc. Transition contexts are contexts of transitions between two adjacent input elements in historical sequences. Specifically, transition contexts denote time intervals between adjacent behaviors. It captures context-adaptive transition effects from past behaviors to future behaviors with different time intervals. Usually, shorter time intervals have more significant effects comparing with longer ones.

In this work, we propose a novel model, named Context-Aware Recurrent Neural Networks (\textbf{CA-RNN}), to model sequential information and contextual information in one framework. Each layer of conventional RNN contains an input element and recurrent transition from the previous status, which are captured by an input matrix and a transition matrix respectively. Different from conventional RNN using a constant input matrix, CA-RNN employs adaptive context-specific input matrices for input elements under various input contexts. Similarly, instead of using a constant transition matrix, CA-RNN utilizes adaptive context-specific transition matrices for modeling varied transition effects from previous input elements under specific transition contexts. Then, we incorporate Bayesian Personalized Ranking (BPR) \cite{rendle2009bpr} and Back Propagation Through Time (BPTT) \cite{rumelhart1988learning} for the learning of CA-RNN. In summary, the main contributions of this work are listed as follows:

\begin{itemize}
\item
We address the problem of context-aware sequential recommendation, which includes two types of contexts, i.e., input contexts and transition contexts.

\item
We employ adaptive context-specific input matrices to model input contexts. These contexts denote external situations under which users conduct behaviors.

\item
We incorporate adaptive context-specific transition matrices for modeling transition contexts, i.e., time intervals between adjacent behaviors in historical sequences.

\item
Experiments conducted on two real-world datasets show that CA-RNN is effective and outperforms state-of-the-art methods significantly.

\end{itemize}

\section{Proposed Model}

In this section, we introduce the proposed Context-Aware Recurrent Neural Networks (CA-RNN). We first present the problem definition and the conventional RNN model. Then we detail our proposed model. Finally, we introduce the parameter learning process of the CA-RNN model.

\subsection{Problem Definition}
We have a set of users denoted as $U  = \{ u_1 ,u_2 ,...\}$, and a set of items denoted as $V  = \{ v_1 ,v_2 ,...\}$. For each user $u$, the behavioral history is given as $V^u = \{v_1^{u}, v_2^{u}, ...\}$, where $v^u_{k}$ denotes the $k$-th selected item of user $u$. Each selected item in the behavior history is associated with corresponding timestamps $T^u = \{t^u_{1}, t^u_{2}, ...\}$, where $t_k^u$ denotes the $k$-th timestamp in the behavioral sequence of user $u$. For each user $u$, at specific timestamp $t_k^u$, the input context is denoted as $c_{I,k}^{u}$, such as weather, location and so on. And the corresponding transition context is denoted as $c_{T,k}^{u}$, which is determined by the time interval between the timestamp $t_k^u$ of the current behavior and the timestamp $t_{k-1}^u$ of the previous behavior.

In this paper, given the behavioral history $V^u = \{v^u_{1},..., v^u_{k},... \}$ of user $u$ with input and transition contexts, we would like to predict the next selected item $v^u_{k+1}$ of user $u$ under input context $c_{I,k+1}^{u}$ and transition context $c_{T,k+1}^{u}$.

\begin{figure*}[htb]
\centering
\setlength{\abovecaptionskip}{0pt}
\setlength{\belowcaptionskip}{-15pt}
\includegraphics[width=1\linewidth]{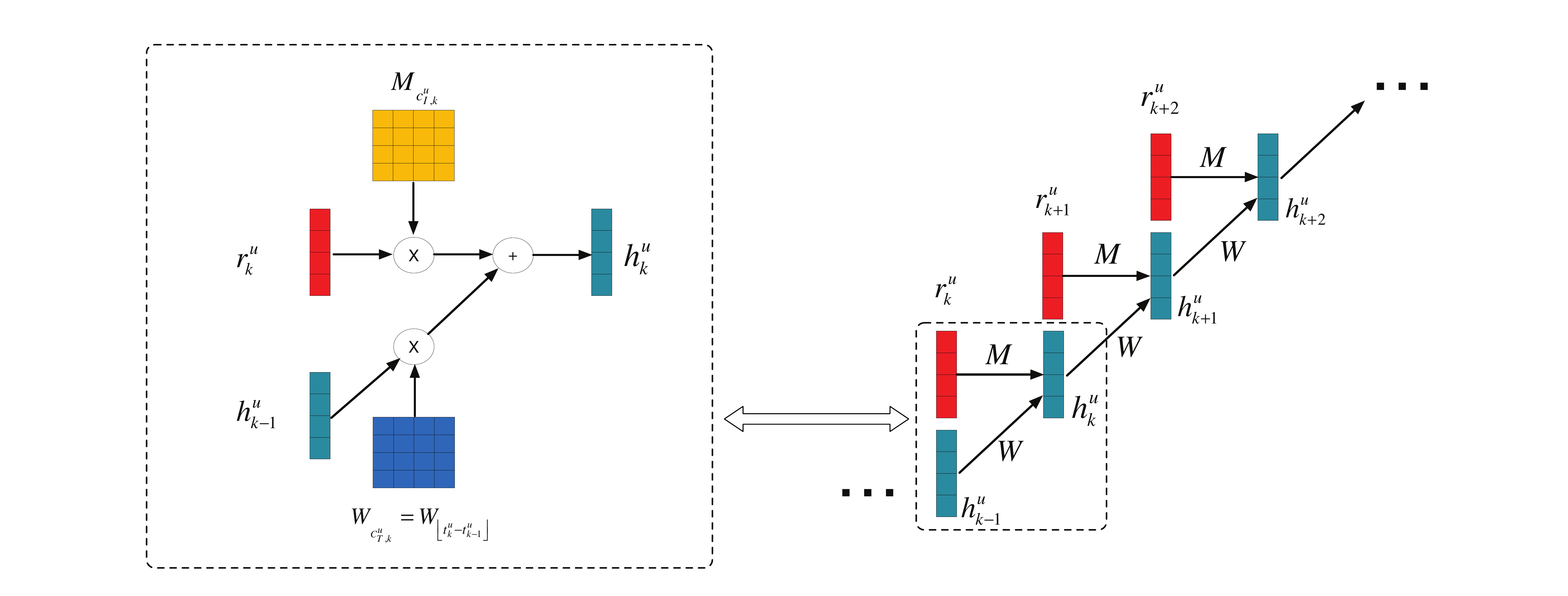}
\caption{Overview of the proposed CA-RNN model. The right part shows the forward propagation process of CA-RNN. The left part illustrates the computational procedure of a hidden layer in CA-RNN.}
\label{fig:Model-overview}
\end{figure*}

\subsection{Recurrent Neural Networks}
The architecture of a recurrent neural networks is a recurrent structure with multiple hidden layers. At each time step, we can predict the output unit given the hidden layer, and then feed the new output back into the next hidden status. The formulation of the hidden layer in RNN is:
\begin{equation}  \label{eqHoriginal}
\mathbf{h}_{k}^{u}=f\left ( {\mathbf{r}_{k}^{u}}\mathbf{M}+\mathbf{h}_{k-1}^{u}\mathbf{W} \right),~
\end{equation}
where $\mathbf{h}^u_{k}$ is the $d$ dimensional hidden status of user $u$ at the timestamp $t_k^u$ in the behavior sequence, and $\mathbf{r}_{k}^{u}$ denotes the $d$ dimensional latent vector of the corresponding item $v_{k}^{u}$. $\mathbf{W}$ is the $d \times d$ dimensional transition matrix, which works as the recurrent connection to propagate sequential signals from the previous status. $\mathbf{M}$ denotes the $d \times d$ dimensional input matrix, which captures the current behavior of the user on input elements. The activation function $f(x)$ is usually chosen as a $sigmoid$ function $f(x) = \exp \left( {1 \mathord{\left/ \right. \kern-\nulldelimiterspace} 1 + e^{ - x} } \right)$.

Though RNN has achieved successful performances in sequential modeling, it has difficulty in modeling a variety of contextual information. With the increasing of contextual information in practical applications, context-aware sequential recommendation becomes an emerging task. As a consequence, it is appropriate to incorporate significant contextual information in sequences into an adapted RNN architecture.

\subsection{Modeling Input Contexts} \label{input}
Input contexts denote the external situations under which users conduct behaviors. This type of contexts in the behavioral history is significant for behavior prediction. For instance, if a man usually takes exercise in the morning, we can predict he will go to the gym in the morning rather than in the evening, even if there are lots of people going to gyms in the evening. Thus, we plan to incorporate input contexts in the conventional RNN model. The constant input matrix in conventional RNN is replaced by adaptive context-specific input matrices according to different input contexts. Then, Equation \ref{eqHoriginal} can be rewritten as:
\begin{equation}\label{eqHt}
\mathbf{h}_{k}^{u}=f\left ( \mathbf{r}_{k}^{u}\mathbf{M}_{c_{I,k}^{u}}+\mathbf{h}_{k-1}^{u}\mathbf{W}\right )~,
\end{equation}
where $\mathbf{M}_{c_{I,k}^{u}}$ denotes the $d \times d$ dimensional context-specific input matrix of input context $c_{I,k}^{u}$. It incorporates external contexts into items so as to help appropriately shape interests of users under specific contexts.

\subsection{Modeling Transition Contexts}
Transition contexts denote lengths of time intervals between adjacent behaviors in historical sequences. Different lengths of time intervals between two behaviors have different impacts on prediction. Generally speaking, longer time intervals usually have weaker effect in predicting next behavior than shorter ones. Therefore, modeling lengths of time interval is essential for predicting future behaviors. Thus, we treat time intervals as transition contexts, and incorporate transition contexts into our model. The constant transition matrix in conventional RNN is replaced with adaptive context-specific transition matrices according to different transition contexts. Then Equation \ref{eqHt} can be rewritten as:
\begin{equation} \label{eqh}
\mathbf{h}_{k}^{u}=f\left ( \mathbf{r}_{k}^{u}\mathbf{M}_{c_{I,k}^{u}}+\mathbf{h}_{k-1}^{u}\mathbf{W}_{c_{T,k}^{u}}\right )~,
\end{equation}
where $\mathbf{W}_{c_{T,k}^{u}}$ is a $d \times d$ dimensional adaptive context-specific transition matrix for the transition context $c_{T,k}^{u}$. It captures how transition contexts between adjacent behaviors affect the transition of global sequential features. The adaptive context-specific transition matrix can be generated according to the length of time interval between the current timestamp and the previous timestamp.

However, it is impossible to learn a context-specific transition matrix for every possible continuous time interval value independently. So we partition the range of all possible time intervals into discrete time bins, and all time intervals are discretized to the floor of the corresponding time bin. Thus, context-specific transition matrices can be generated as:
\begin{equation} \label{genge}
\mathbf{W}_{c_{T,k}^{u}} = \mathbf{W}_{\left\lfloor t_k^u-t_{k-1}^u \right\rfloor},
\end{equation}
where $t_k^u$ and $t_{k-1}^u$ denotes the current and the previous timestamp respectively, $t_k^u-t_{k-1}^u$ is the length of the corresponding time interval, and $\left\lfloor t_k^u-t_{k-1}^u \right\rfloor$ denotes the floor of the length of corresponding time interval $t_k^u-t_{k-1}^u$.

Then, as shown in Figure \ref{fig:Model-overview}, the formulation of each hidden layer in the CA-RNN model becomes:
\begin{equation} \label{eqh3}
\mathbf{h}_{k}^{u}=f\left (\mathbf{r}_{k}^{u}\mathbf{M}_{c_{I,k}^{u}}+\mathbf{h}_{k-1}^{u}\mathbf{W}_{\left\lfloor t_k^u-t_{k-1}^u \right\rfloor}\right )~.
\end{equation}

\subsection{Context-aware Prediction}

When making prediction of user behaviors at timestamp $t_{k+1}^u$, despite historical behavioral contexts modeled above, current contexts $c_{I,k+1}^{u}$ and $c_{T,k+1}^{u}$ also have significant effects. Thus, our method should incorporate current contexts to make context-aware prediction. The prediction function, i.e., whether user $u$ will select item $v$ at timestamp $t_{k+1}^u$ under contexts $c_{I,k+1}^{u}$ and $c_{T,k+1}^{u}$, can be written as:
\begin{equation} \label{prediction}
y_{u, k+1, v} = \mathbf{h}_{k}^{u} \mathbf{W'}_{c_{T,k+1}^{u}} \left( \mathbf{r}_{v} \mathbf{M'}_{c_{I,k+1}^{u}} \right) ^ \mathrm{ T }~,
\end{equation}
where $\mathbf{M'}_{c_{I,k+1}^{u}}$ and $\mathbf{W'}_{c_{T,k+1}^{u}}$ denote $d \times d$ dimensional matrix representations of the current input context $c_{I,k+1}^{u}$ and transition context $c_{T,k+1}^{u}$ respectively. $\mathbf{W'}_{c_{T,k+1}^{u}}$ can be generated according to the calculation in Equation \ref{genge}.

\subsection{Parameter Learning}

In this subsection, we introduce the learning process of CA-RNN with BPR \cite{rendle2009bpr} and BPTT \cite{rumelhart1988learning}. These methods have been successfully used for parameter estimation of RNN based recommendation models \cite{liu2016strnn}\cite{yu2016dream}.

BPR is a widely-used pairwise ranking framework for the implicit feedback data. The basic assumption of BPR is that a user prefers selected items than negative ones. Under the BPR framework, at each sequential position $k$, the objective of CA-RNN is to maximize the following probability:
\begin{equation}
p(u, k, v\succ v') = g(y_{u, k, v} - y_{u, k, v'}),~
\end{equation}
where $v'$ denotes a negative item sample and $g(x)$ is a nonlinear function which can be defined as $g(x) = {1 \mathord{\left/ \right. \kern-\nulldelimiterspace} 1 + e^{ - x}}$. Incorporating the negative log likelihood, we can solve the following objective function equivalently:
\begin{equation}
J = \sum\limits_{u,k} \ln(1+e^{-(y_{u,k,v} - y_{u,k,v'})}) +\frac{\lambda }{2}\left \| \mathbf{\Theta}  \right \|^{2}  ~,
\end{equation}
where $\mathbf{\Theta} = \{\mathbf{R}, \mathbf{M} ,\mathbf{W}\}$ denotes all the parameters to be estimated, and $\lambda$ is a parameter to control the power of regularization. According to Equation \ref{prediction}, derivations of $J$ with respect to the parameters can be calculated. Moreover, parameters in CA-RNN can be further learnt by using the BPTT algorithm. According to Equation \ref{eqh}, given the derivation ${\partial J \mathord{\left/ \right. \kern-\nulldelimiterspace} {\partial \mathbf{h}_{k}^{u} } }$, corresponding gradients of all parameters in the hidden layer can be calculated, which is repeated iteratively in the whole sequence. And stochastic gradient descent can be employed to estimate all the parameters.

\begin{table*}[htbp]
\centering
\caption{Performance comparison between input contexts and transition contexts on the Taobao and Movielens-1M datasets with dimensionality $d=10$ evaluated by Recall, F1-score, MAP and NDCG.}
    \begin{tabular}{cccccccccc}
    \toprule
    Dataset & Method & Recall@1 & Recall@5 & Recall@10 & F1-score@1 & F1-score@5 & F1-score@10 & MAP   & NDCG \\
    \midrule
    \multirow{3}[0]{*}{Taobao} & CA-RNN-input & 0.1206  & 0.3376  & 0.4914  & 0.1206  & 0.1125  & 0.0893  & 0.2346  & 0.3684  \\
          & CA-RNN-transition & 0.1129  & 0.3284  & 0.4392  & 0.1129  & 0.1095  & 0.0799  & 0.2186  & 0.3500  \\
          & CA-RNN & \textbf{0.1648} & \textbf{0.4305} & \textbf{0.5804} & \textbf{0.1648} & \textbf{0.1435} & \textbf{0.1055} & \textbf{0.2932} & \textbf{0.4190} \\
    \midrule
    \multirow{3}[0]{*}{Movielens-1M} & CA-RNN-input & 0.0070  & 0.0337  & 0.0521  & 0.0070  & 0.0112  & 0.0095  & 0.0400  & 0.1478  \\
          & CA-RNN-transition & 0.0071  & 0.0376  & 0.0536  & 0.0071  & 0.0125  & 0.0097  & 0.0407  & 0.1484  \\
          & CA-RNN & \textbf{0.0075} & \textbf{0.0394} & \textbf{0.0551} & \textbf{0.0075} & \textbf{0.0131} & \textbf{0.0100} & \textbf{0.0412} & \textbf{0.1493} \\
    \bottomrule
    \end{tabular}%
\label{tab:InputAndTransition}%
\end{table*}%

\section{Experiments}
In this section, we conduct empirical experiments to demonstrate the effectiveness of CA-RNN on context-aware sequential recommendation. First, we introduce our experimental settings. Then we conduct experiments on comparing input contexts and transition contexts. We also compare our proposed CA-RNN model with some state-of-the-art sequential and context-aware recommendation methods. Finally, we study the impact of dimensionality in CA-RNN.

\subsection{Experimental Settings}
Our experiments are conducted on two real datasets with rich contextual information. The \textbf{Taobao} dataset\footnote{https://tianchi.shuju.aliyun.com/competition/index.htm} is a public online e-commerce dataset collected from taobao.com\footnote{https://www.taobao.com/}. It contains 13,611,038 shopping records belonging to 1,103,702 users and 462,008 items. The \textbf{Movielens-1M} dataset\footnote{http://files.grouplens.org/datasets/movielens/ml-1m.zip} is a dataset for recommender systems. It contains 1,000,209 rating records of 3,900 movies by 6,040 users. Records in both datasets are associated with timestamps.

Moreover, for each behavioral sequence in the two datasets, we use first $80\%$ elements in the sequence for training, the remaining $20\%$ data for testing. And the regulation parameter in our experiments is set as $\lambda = 0.01$. To avoid data sparsity, on both datasets, we remove users with less than 10 records and items with less than 3 records.

According to the contextual information in the two datasets, we extract input contexts and transition contexts to implement our proposed CA-RNN model, as well as other context-aware methods. First, based on timestamps, we can extract different kinds of input contexts on the two datasets. On the Taobao dataset, we extract three kinds of input contexts: seven days in a week, three ten-day time periods in a month, and holiday or not. So, there are totally 42 input context values in the Taobao dataset. On the Movielens-1M dataset, we extract two kinds of input contexts: seven days in a week, and twenty-four hours in a day. So, there are totally 168 input context values in the Movielens-1M dataset. Second, we can extract time intervals between adjacent behaviors in sequences as transition contexts. Time intervals in both datsets are discretized to one-day time bins. To avoid data sparsity, for time intervals larger than 30 days, we treat them as one time bin. So, there are totally 32 transition context values in both datasets.

Moreover, we evaluate performances of each method through several widely-used evaluation metrics for ranking tasks: \textbf{Recall}@k, \textbf{F1-score}@k, Mean Average Precision (\textbf{MAP}) and Normalized Discounted Cumulative Gain (\textbf{NDCG}). For Recall@k and F1-score@k, we report results with $k=1$, $5$ and $10$ in our experiments. For all the metrics, the larger the value, the better the performance.

\begin{table*}[htbp]
\centering
\caption{Performance comparison of CA-RNN and other compared methods on the Taobao and Movielens-1M datasets with dimensionality $d=10$ evaluated by Recall, F1-score, MAP and NDCG.}
    \begin{tabular}{cccccccccc}
    \toprule
    Dataset & Method & Recall@1 & Recall@5 & Recall@10 & F1-score@1 & F1-score@5 & F1-score@10 & MAP   & NDCG \\
    \midrule
    \multirow{8}[0]{*}{Taobao} & POP   & 0.0024  & 0.0246  & 0.0272  & 0.0024  & 0.0082  & 0.0049  & 0.0314  & 0.1290  \\
          & BPR   & 0.0145  & 0.0462  & 0.0783  & 0.0145  & 0.0154  & 0.0142  & 0.0510  & 0.1678  \\
          & FM    & 0.0241  & 0.0660  & 0.1070  & 0.0241  & 0.0220  & 0.0195  & 0.0732  & 0.2119  \\
          & CARS2 & 0.0257  & 0.0687  & 0.1107  & 0.0257  & 0.0229  & 0.0201  & 0.0761  & 0.2166  \\
          & FPMC  & 0.0581  & 0.1869  & 0.2776  & 0.0581  & 0.0623  & 0.0505  & 0.1261  & 0.2544  \\
          & HRM   & 0.0629  & 0.2024  & 0.2969  & 0.0629  & 0.0675  & 0.0540  & 0.1366  & 0.2672  \\
          & RNN   & 0.0859  & 0.2696  & 0.3849  & 0.0859  & 0.0899  & 0.0700  & 0.1828  & 0.3186  \\
          & CA-RNN & \textbf{0.1648} & \textbf{0.4305} & \textbf{0.5804} & \textbf{0.1648} & \textbf{0.1435} & \textbf{0.1055} & \textbf{0.2932} & \textbf{0.4190} \\
    \midrule
    \multirow{8}[0]{*}{Movielens-1M} & POP   & 0.0016  & 0.0052  & 0.0118  & 0.0016  & 0.0017  & 0.0021  & 0.0096  & 0.0434  \\
          & BPR   & 0.0043  & 0.0237  & 0.0436  & 0.0043  & 0.0079  & 0.0079  & 0.0302  & 0.1188  \\
          & FM    & 0.0044  & 0.0246  & 0.0446  & 0.0044  & 0.0082  & 0.0081  & 0.0312  & 0.1229  \\
          & CARS2 & 0.0045  & 0.0249  & 0.0448  & 0.0045  & 0.0083  & 0.0081  & 0.0314  & 0.1234  \\
          & FPMC  & 0.0054  & 0.0255  & 0.0440  & 0.0054  & 0.0085  & 0.0080  & 0.0332  & 0.1291  \\
          & HRM   & 0.0060  & 0.0288  & 0.0462  & 0.0060  & 0.0096  & 0.0084  & 0.0345  & 0.1318  \\
          & RNN   & 0.0063  & 0.0318  & 0.0484  & 0.0063  & 0.0106  & 0.0088  & 0.0362  & 0.1364  \\
          & CA-RNN & \textbf{0.0075} & \textbf{0.0394} & \textbf{0.0551} & \textbf{0.0075} & \textbf{0.0131} & \textbf{0.0100} & \textbf{0.0412} & \textbf{0.1493} \\
    \bottomrule
    \end{tabular}%
\label{tab:result}%
\end{table*}%

\begin{figure*}[tb]
\centering
\setlength{\abovecaptionskip}{0pt}
\setlength{\belowcaptionskip}{-15pt}
\subfigure[Performances on the Taobao dataset.]{
\begin{minipage}[b]{0.48\textwidth}
\includegraphics[width=1\textwidth]{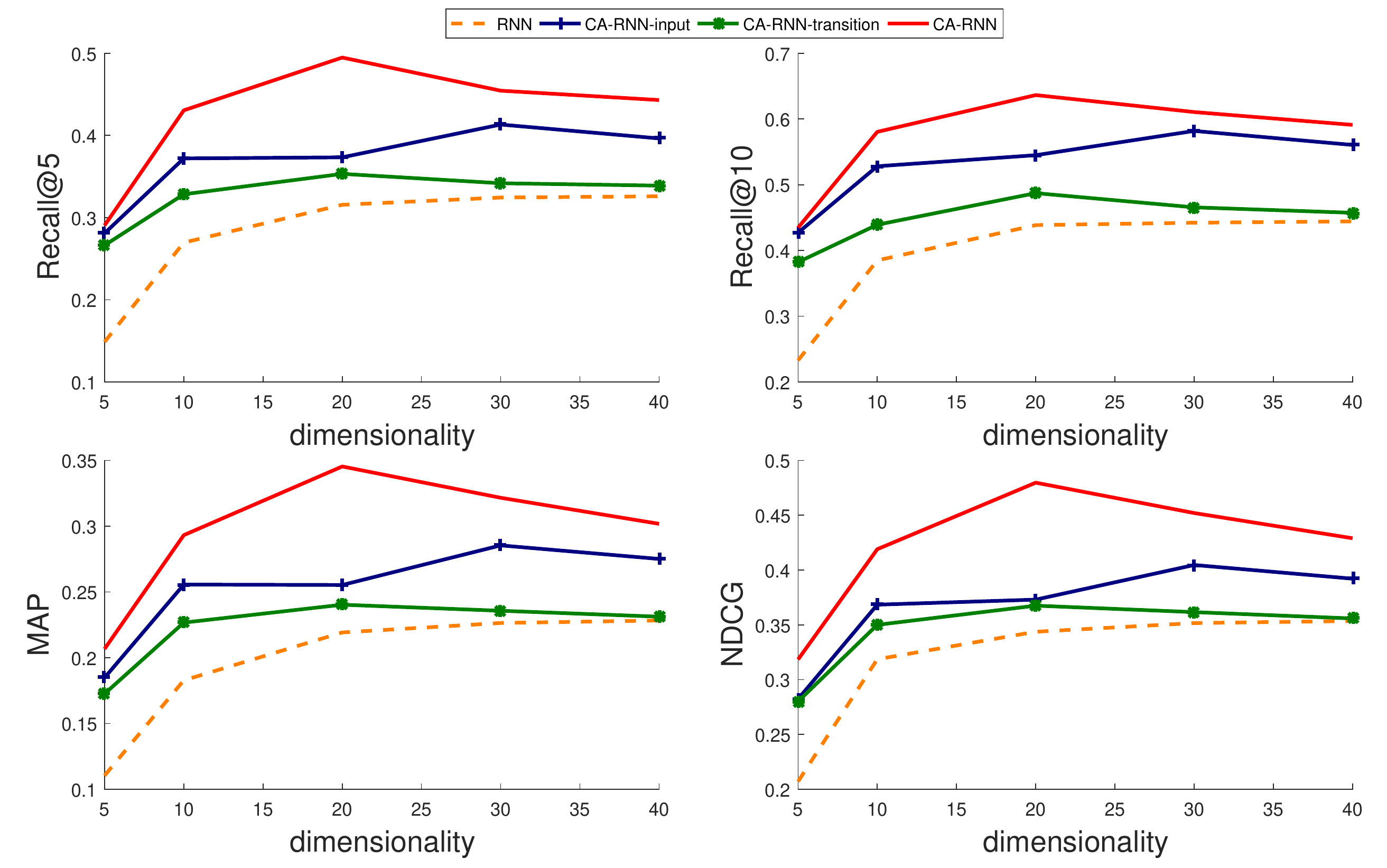}
\label{dim_taobao}
\end{minipage}
}
\vspace{-1mm}
\subfigure[Performances on the Movielens-1M dataset.]{
\begin{minipage}[b]{0.48\textwidth}
\includegraphics[width=1\textwidth]{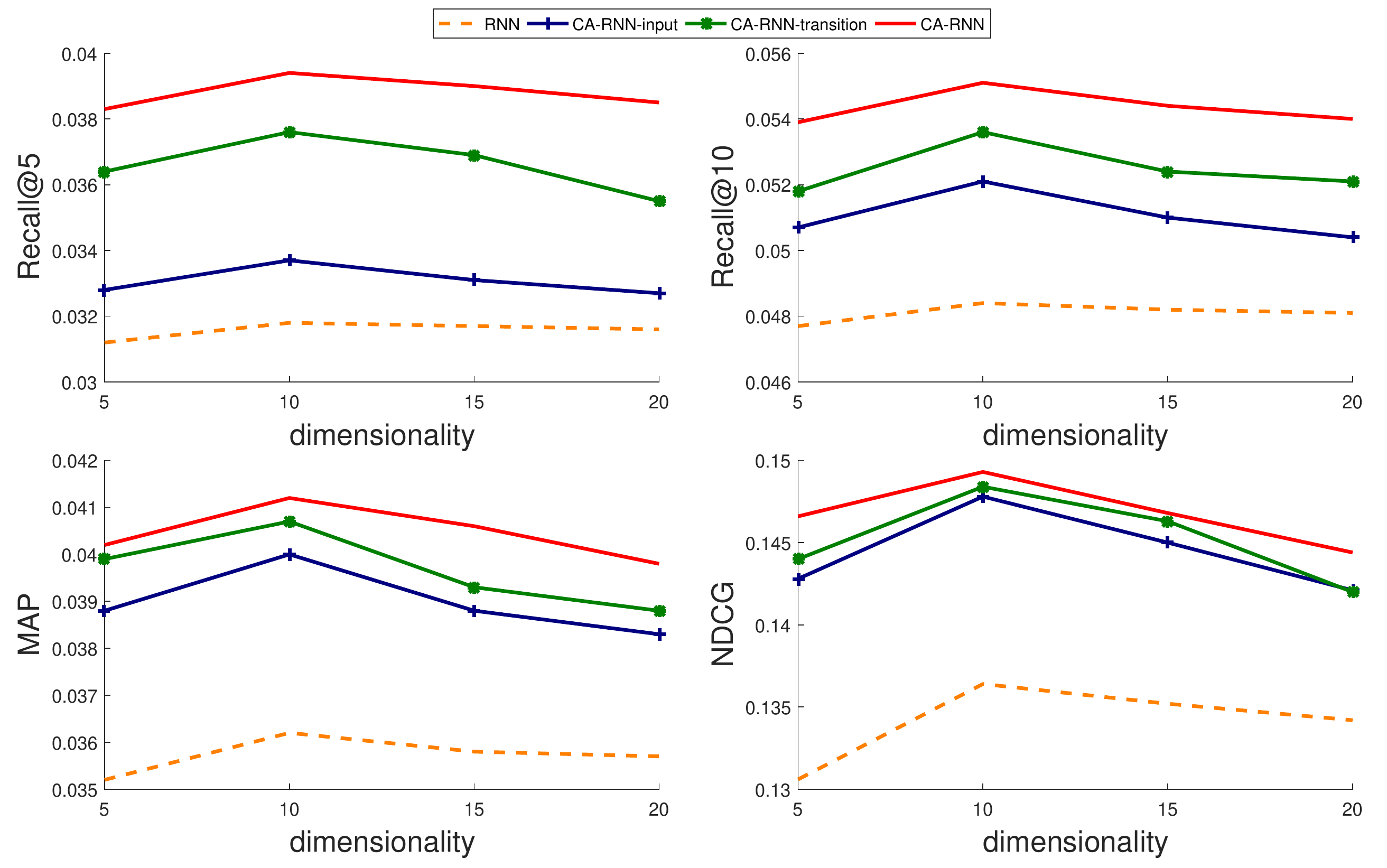}
\label{dim_movie}
\end{minipage}
}
\caption{Performances of RNN, CA-RNN-input, CA-RNN-transition and CA-RNN with varying dimensionality $d$.}
\label{fig:dim}
\end{figure*}

In our experiments, to investigate the effectiveness of CA-RNN from various angles, there are several aspects of compared methods: baseline methods, context-aware methods, sequential methods and our proposed methods. Baseline methods contain \textbf{POP} and \textbf{BPR} \cite{rendle2009bpr}. Context-aware methods consist of \textbf{FM} \cite{rendle2011fast} and \textbf{CARS2} \cite{shi2014cars}. And sequential methods include \textbf{FPMC} \cite{rendle2010factorizing}, \textbf{HRM} \cite{wang2015learning} and \textbf{RNN} \cite{yu2016dream}. The length of each transaction in FPMC and HRM is set to be one day and one week on the Taobao and Movielens-1M datasets respectively. Moreover, to compare the effects of input contexts and transition contexts, we implement not only the \textbf{CA-RNN} model\footnote{Code: \url{http://qiangliucasia.github.io/homepage/files/CARNNcode.zip}}, but also \textbf{CA-RNN-input} and \textbf{CA-RNN-transition}, with only input or transition contexts.

\subsection{Input Contexts VS. Transition Contexts}
To compare input contexts and transition contexts, we illustrate performances of CA-RNN-input, CA-RNN-transition and CA-RNN in Table \ref{tab:InputAndTransition}. On the Taobao dataset, CA-RNN-input outperforms CA-RNN-transition. While on the Movielens-1M dataset, CA-RNN-transition outperforms CA-RNN-input. These results indicate that the contextual information in input contexts and transition contexts has distinct effects on predicting future behaviors, and their significance differs in different situations. Thus, it is necessary to incorporate both types of contexts to better predict the future. Incorporating both input contexts and transition contexts, CA-RNN achieves the best performance among our proposed methods, and can significantly outperform CA-RNN-input and CA-RNN-transition.

\subsection{Performance Comparison}
To evaluate the effectiveness of our proposed CA-RNN, we illustrate the performance comparison of CA-RNN and other compared methods in Figure \ref{tab:result}. Comparing with baseline performances of POP and BPR, context-aware methods FM and CARS2 achieve significant improvements. And sequential methods FPMC, HRM and RNN can further outperform FM and CARS2. This indicates that, comparing with contextual information, sequential information has more significant effects. And RNN achieves the best performance among all the compared methods. Moreover, we can observe that, our proposed CA-RNN greatly outperforms compared methods on both datasets in terms of all the metrics. Comparing with RNN, CA-RNN relatively improves Recall@1, @5, @10, MAP and NDCG by $91.8\%$, $59.7\%$, $50.8\%$, $60.4\%$ and $31.5\%$ on the Taobao dataset. And on the Movielens-1M datasets, the relative improvements become $19.1\%$, $23.9\%$, $13.9\%$, $13.8\%$ and $9.5\%$. These improvements indicate the superiority of our method brought by modeling input and transition contexts.

\subsection{Impact of Dimensionality}
To study the impact of dimensionality in our methods, we illustrate performances of RNN-based methods with varying $d$ in Figure \ref{fig:dim}. On the Taobao dataset, the best performances of CA-RNN, CA-RNN-transition and CA-RNN-input are achieved at $d=20$, $d=20$ and $d=30$ respectively. On Movielens-1M, performances of our methods are stable along with the dimensionality, and best performances are achieved at $d=10$. On both dataset, our methods can stably outperform conventional RNN with a significant advantage in all situations. Moreover, even not with the best dimensionality, CA-RNN can still outperform conventional RNN.

\section{Conclusions}
In this paper, we address the problem of context-aware sequential recommendation and propose a novel method, i.e., context-aware recurrent neural networks. In CA-RNN, the constant input matrix of conventional RNN is replaced with context-specific input matrices to modeling complex real-world contexts, e.g., time, location and weather. Meanwhile, to model transition contexts, i.e., time intervals between adjacent behaviors in historical sequences, context-specific transition matrices are incorporated, instead of the constant one in conventional RNN. The experimental results on two real datasets show that CA-RNN outperforms competitive sequential and context-aware models.

\section*{Acknowledgments}
This work is jointly supported by National Basic Research Program of China (2012CB316300), and National Natural Science Foundation of China (61403390, U1435221, 61420106015, 61525306).

\bibliographystyle{abbrv}
\bibliography{IEEE}

\end{document}